\documentclass[12pt, a4paper]{article}

\usepackage{amsmath} 
\usepackage{amssymb} 

\usepackage[margin=0.9in]{geometry} 
\usepackage[parfill]{parskip} 
\usepackage{setspace} 
\usepackage{titlesec} 
\titleformat*{\section}{\large\bfseries}
\titleformat*{\subsection}{\normalsize\bfseries}
\titleformat*{\subsubsection}{\normalsize\bfseries}
\titleformat*{\paragraph}{\normalsize\bfseries}
\titleformat*{\subparagraph}{\normalsize\bfseries}
\usepackage{titling} 

\usepackage{graphicx} 
\usepackage{float} 
\usepackage[font=footnotesize,labelfont=bf]{caption} 

\usepackage{hyperref}
\hypersetup{colorlinks, linkcolor=purple, citecolor=cyan, urlcolor=blue}

\usepackage{authblk} 
\usepackage{orcidlink}

\usepackage[style=nature]{biblatex}
\renewcommand{\cite}{\supercite} 
\begin{document}

\title{Personalized identification, prediction, and stimulation of neural oscillations via data-driven models of epileptic network dynamics}

\date{}

\author[1,2]{Tena Dubcek\orcidlink{0000-0003-3636-3201}\thanks{tena.dubcek@kliniklengg.ch / dubcekt@ethz.ch}}
\author[1]{Debora Ledergerber}
\author[2]{Jana Thomann}
\author[2]{Giovanna Aiello}
\author[3]{Marc Serra Garcia}
\author[$\dagger$1]{Lukas Imbach}
\author[$\dagger$2]{Rafael Polania}

\affil[1]{\footnotesize Swiss Epilepsy Center, Klinik Lengg, Zurich, Switzerland}
\affil[2]{\footnotesize ETH Zurich, Department of Health Sciences and Technology, Switzerland}
\affil[3]{\footnotesize AMOLF, Amsterdam, Netherlands}
\affil[$\dagger$]{\footnotesize\textit{These authors contributed equally to this work.}} 

\maketitle

\begin{abstract}
Neural oscillations are considered to be brain-specific signatures of information processing and communication in the brain. They also reflect pathological brain activity in neurological disorders, thus offering a basis for diagnoses and forecasting. Epilepsy is one of the most common neurological disorders, characterized by abnormal synchronization and desynchronization of the oscillations in the brain. About one third of epilepsy cases are pharmacoresistant, and as such emphasize the need for novel therapy approaches, where brain stimulation appears to be a promising therapeutic option. The development of brain stimulation paradigms, however, is often based on generalized assumptions about brain dynamics, although it is known that significant differences occur between patients and brain states.
We developed a framework to extract individualized predictive models of epileptic network dynamics directly from EEG data. The models are based on the dominant coherent oscillations and their dynamical coupling, thus combining an established interpretation of dynamics through neural oscillations, with accurate patient-specific features. We show that it is possible to build a direct correspondence between the models of brain-network dynamics under periodic driving, and the mechanism of neural entrainment via periodic stimulation. When our framework is applied to EEG recordings of patients in status epilepticus\textemdash a brain state of perpetual seizure activity, it yields a model-driven predictive analysis of the therapeutic performance of periodic brain stimulation. This suggests that periodic brain stimulation can drive pathological states of epileptic network dynamics towards a healthy functional brain state. 
\end{abstract}


\section{Introduction}
\label{Introduction}

Epilepsy is one of the most common neurological disorders, affecting $50$ million people worldwide~\cite{devinskyEpilepsy2018}. Strikingly, about one third of epilepsy cases are pharmacoresistant~\cite{devinskyEpilepsy2018}, which leads to uncontrolled seizures and a poor quality of life.
Non-pharmacological approaches, such as epilepsy surgery~\cite{zijlmansChangingConceptsPresurgical2019} or brain stimulation~\cite{ryvlinNeuromodulationEpilepsyStateoftheart2021}, emphasize the need for a good understanding of the dynamics in the underlying epileptic network~\cite{kramerEpilepsyDisorderCortical2012a, liNeuralFragilityEEG2021}, and how these dynamics can be altered when they are pathological. Therapies by brain stimulation~\cite{ryvlinNeuromodulationEpilepsyStateoftheart2021, xueNeuromodulationDrugresistantEpilepsy2022} attempt to modulate collective neural dynamics, ideally by targeting the critical nodes in the epileptic network and the relevant oscillatory patterns. The details of stimulation paradigms, however, are often based on generalized assumptions that may disregard the specific spatial and temporal properties of a certain brain network. These brain networks, on the contrary, exhibit immense heterogeneity, especially in epilepsy patients. Therefore, in order to attain the best performance, the personalized angle of brain stimulation therapies and the neural oscillations they target has to be strengthened.

Neural oscillations are essential for information processing and communication in the brain, and patient-specific changes in oscillatory activity provide valuable insights into the state and progression of pathological brain activity~\cite{siegelSpectralFingerprintsLargescale2012a, friesRhythmsCognitionCommunication2015, buzsakiScalingBrainSize2013}. Synchronization of neurons occurs (i) locally, where it enables to cohesively incorporate meaningful information, and (ii) at larger distances, where it ensures the temporal coordination of presynaptic and postsynaptic activation patterns in brain networks~\cite{friesRhythmsCognitionCommunication2015}. Despite the clear relevance of synchronization and coherence, conclusions and predictions about brain rhythmic dynamics are often drawn from measures that obscure these properties. For example, the most common approach used to determine individual alpha rhythms consists in finding the peak or "center-of-gravity" of the power spectral density in the {\it a priori} defined alpha frequency band~\cite{corcoranReliableAutomatedMethod2018a}. Power spectral density, however, does not distinguish the consistently present coherent oscillations, which can carry meaningful information, from accidental noisy contributions, which most probably are not interpretable in the context of brain processes. Such an approach is particularly problematic when combined with dynamics in which the processes of (de)synchronization and entrainment are accentuated. The dynamics of an epileptic brain network, with their distinctive abnormal synchronization and desynchronization episodes~\cite{jiruskaSynchronizationDesynchronizationEpilepsy2013a}, certainly stand within this domain. For example, EEG recordings in status epilepticus might reflect highly coherent epileptic network oscillations and local phenomena of EEG synchronization simultaneously. When these dynamics are modulated by brain stimulation, it becomes even more important to correctly identify and understand the relevant coherent oscillations.

Recent evidence shows that neural oscillations can be modulated by directly stimulating chosen neuronal populations via weak periodically varying currents, such as in transcranial alternating current stimulation (tACS)~\cite{riddleTargetingNeuralOscillations2021, johnsonDosedependentEffectsTranscranial2020, krauseTranscranialElectricalStimulation2023a} or deep brain stimulation (DBS)~\cite{lozanoDeepBrainStimulation2019}. The long-term modulation of the underlying network dynamics relies on synaptic plasticity and the subsequent network reorganization~\cite{herringtonMechanismsDeepBrain2016a, krauseBrainStimulationCompetes2022, vogetiEntrainmentSpikeTimingDependent2022}. These synaptic changes are thought to be dependent on the timings of neuronal firings in the network, induced by the acute response of neurons to periodic stimulation. The acute response of the neural dynamics can be explained primarily by the entrainment theory~\cite{aliTranscranialAlternatingCurrent2013, vogetiEntrainmentSpikeTimingDependent2022}: intrinsic neural oscillations are modulated to become temporally aligned with the periodic stimulation, with changes that depend on the stimulation frequency and location. 
Transcranial alternating current stimulation (tACS), is a stimulation paradigm that appears to safely induce neural entrainment by periodic stimulation in a non-invasive manner\cite{johnsonDosedependentEffectsTranscranial2020, 
beliaevaIntegrativeApproachesStudy2021,
krauseBrainStimulationCompetes2022}. It offers substantial flexibility in the choice of stimulation frequency and location. Therefore, tACS emerges as a promising candidate for interventions in cases of unexpected acute pharmacoresistant epileptic episodes, for which there is not enough time nor a close reasoning to implant the more established invasive stimulation paradigms~\cite{ryvlinNeuromodulationEpilepsyStateoftheart2021b}. Applying tACS in epilepsy, however, is hindered by the incomplete understanding of the mechanisms through which it interacts with brain dynamics, and a reliable procedure that can predict the consequences of its application for each patient.

The most severe brain state of pharmacoresistant epilepsy is refractory status epilepticus\textemdash a brain state of perpetual seizure activity~\cite{trinkaDefinitionClassificationStatus2015, pintoStatusEpilepticusReview2022}. Status epilepticus is a medical emergency and, when refractory, has a mortality rate of $40\%$~\cite{pintoStatusEpilepticusReview2022, trinkaMortalityLifeExpectancy2023}, thus underscoring the need for novel therapy approaches. Remarkably, status epilepticus is a unique epileptic phenomenon also from a dynamical point of view: unlike most other epileptic seizures, which are characterized by a quick progression of dissimilar states of the epileptic network dynamics~\cite{kramerEpilepsyDisorderCortical2012a}, the brain activity in status epilepticus can be well described as steady-state dynamics~\cite{burmanTransitionStatusEpilepticus2020}. Dynamics that are steady-state can be characterized by quantities that are relatively stable in time and the equations describing them are often more interpretable. Therefore, status epilepticus is an excellent point of departure for the development of data-driven predictive models that could help the progress of emerging therapeutic procedures. Namely, once the dynamics in the (pathological) steady-state are understood and captured by a model, an immediate model-based search for the external influence (stimulation) that restores the dynamics of another (healthy) steady state becomes possible. 

Here, we present a procedure to extract personalized interpretable generative models of the steady-state epileptic-network dynamics. In a data-driven and assumption-free manner, we obtain personalized oscillator-network models directly from EEG data. At the same time, by identifying the dominant coherent neural oscillations, the modeling procedure is interpretable within the standard framework and understanding of brain dynamics. Each dynamical model yields an effective dynamical connectivity matrix of the individual network properties, which can predict the network's dynamical reaction to external stimuli. Crucially, we demonstrate that this is useful to understand the patient-specific modulation of neural oscillations via periodic stimulation, and anticipate the most promising patient-specific stimulation parameters that could help the resolution of status epilepticus.


\begin{figure}[H]
\centering
\includegraphics[width=\linewidth]{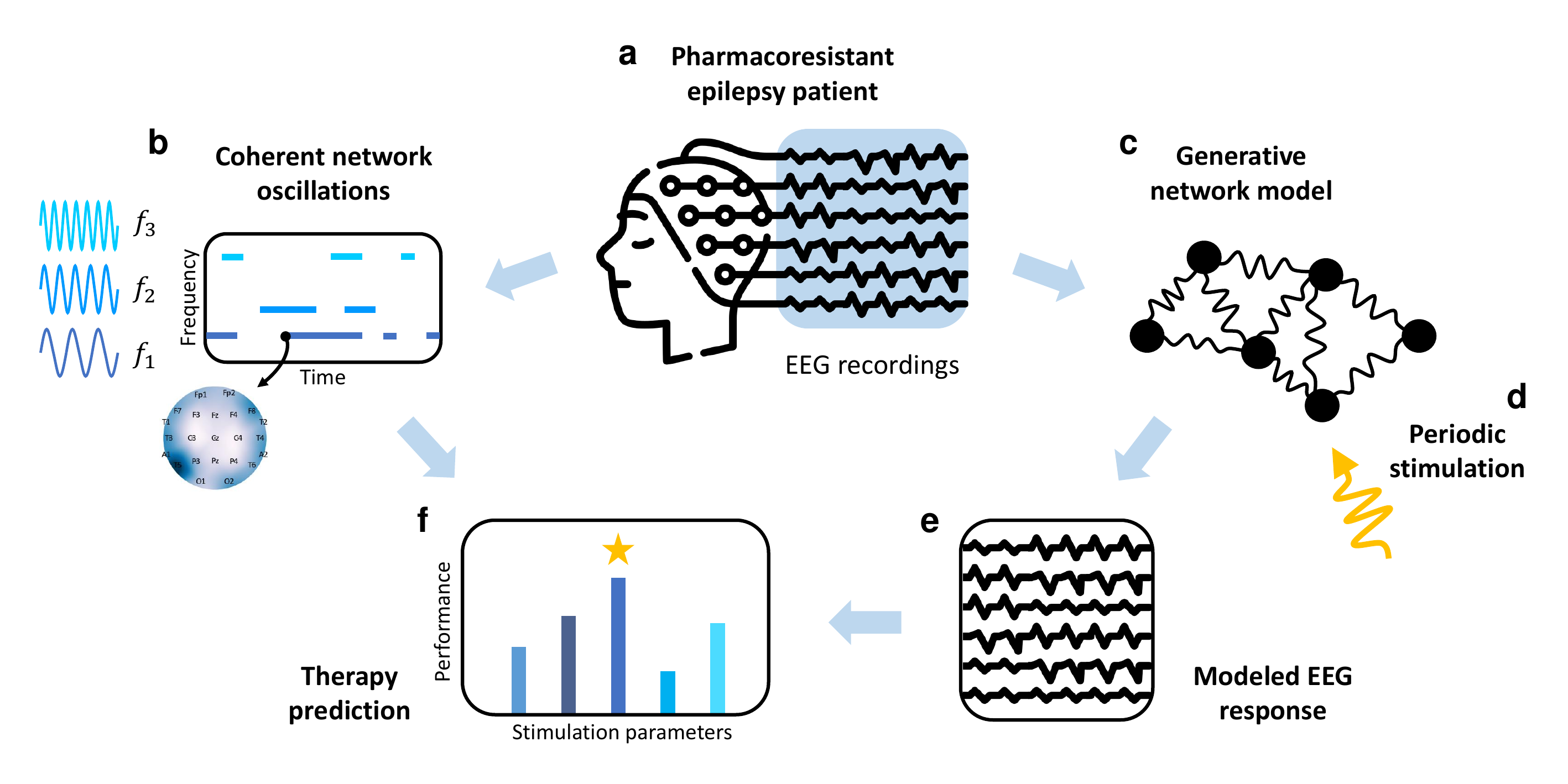}
\caption{
Personalized EEG-driven modeling of epileptic network dynamics and their modulation~|
\textbf{a}~EEG recordings of pharmacoresistant pathological brain activity are used to construct \textbf{b},~\textbf{c}~generative models of the underlying epileptic network dynamics. The interpretable personalized oscillator-network models are extracted \textbf{d} by identifying the dominant coherent oscillations in the EEG dynamics. \textbf{e} The effective dynamical connectivity matrix of the epileptic network is encoded in the models and can be used to predict the network dynamical reaction to external stimuli. \textbf{f} The proposed fully data-drive paradigm can help understand and anticipate the response to available periodic brain stimulation paradigms. 
}
\label{Fig0}
\end{figure}

\section{Results}
\label{Results}

The understanding of ways in which pathological epileptic-network dynamics can be best modulated by external stimulation is still incomplete, although such paradigms are among the most promising solutions in the case of pharmacoresistant epilepsies. Theories of epileptic network dynamics are centered around neural oscillations, mostly explaining the dominant mechanisms through which these oscillations are influenced and entrained by brain stimulation. Yet, in most cases, these theories rely on generalized assumptions about the brain, and thus fail to grasp the patient-specific features of the dynamics\textemdash which are most likely pronounced in patients with pathological brain activity.

EEG provides a window into the scalp signatures of neural oscillations, but these signatures are nontrivial to identify and even harder to directly interpret in a personalized manner. Standard EEG analyses are able to quantify the potential changes to a large extent (e.g., through statistical analyses of changes in spectral and coherence properties), but do not necessarily give a clear pathway to an intuition behind the responsible dynamical mechanism. There remains a missing bridge between the accurate but complex information that is drawn from EEG recordings, and the intuitive theories of brain oscillations and the possible ways to influence them\textemdash in this work, we aimed to establish this missing bridge. The procedure to construct EEG-driven oscillator-network models, which are simultaneously personalized, interpretable and generative, allowed us to combine both sides. As a use case, we analyzed an EEG dataset of $10$ patients in a state of non-convulsive status epilepticus (state name: Status), which was then pharmacologically resolved in $5$ patients (state name: Resolved). The dataset was chosen because it contains both pathological and healthy dynamics happening in the same brain, which makes possible a unique brain-specific comparison of the two states that is not obfuscated by the inter-brain differences.
For each patient, we extracted representative recordings of equal duration for each of the two states. These recordings were used to extract oscillator-network models of the corresponding steady-state dynamics, develop an intuition of the possible transitions under periodic stimulation, and inform about the most promising stimulation parameters (frequency, location) that could lead to status resolution.

\subsection{Identification of robust coherent oscillations}
\label{SS_coherent_oscillations}

\begin{figure}
\centering
\includegraphics[width=\linewidth]{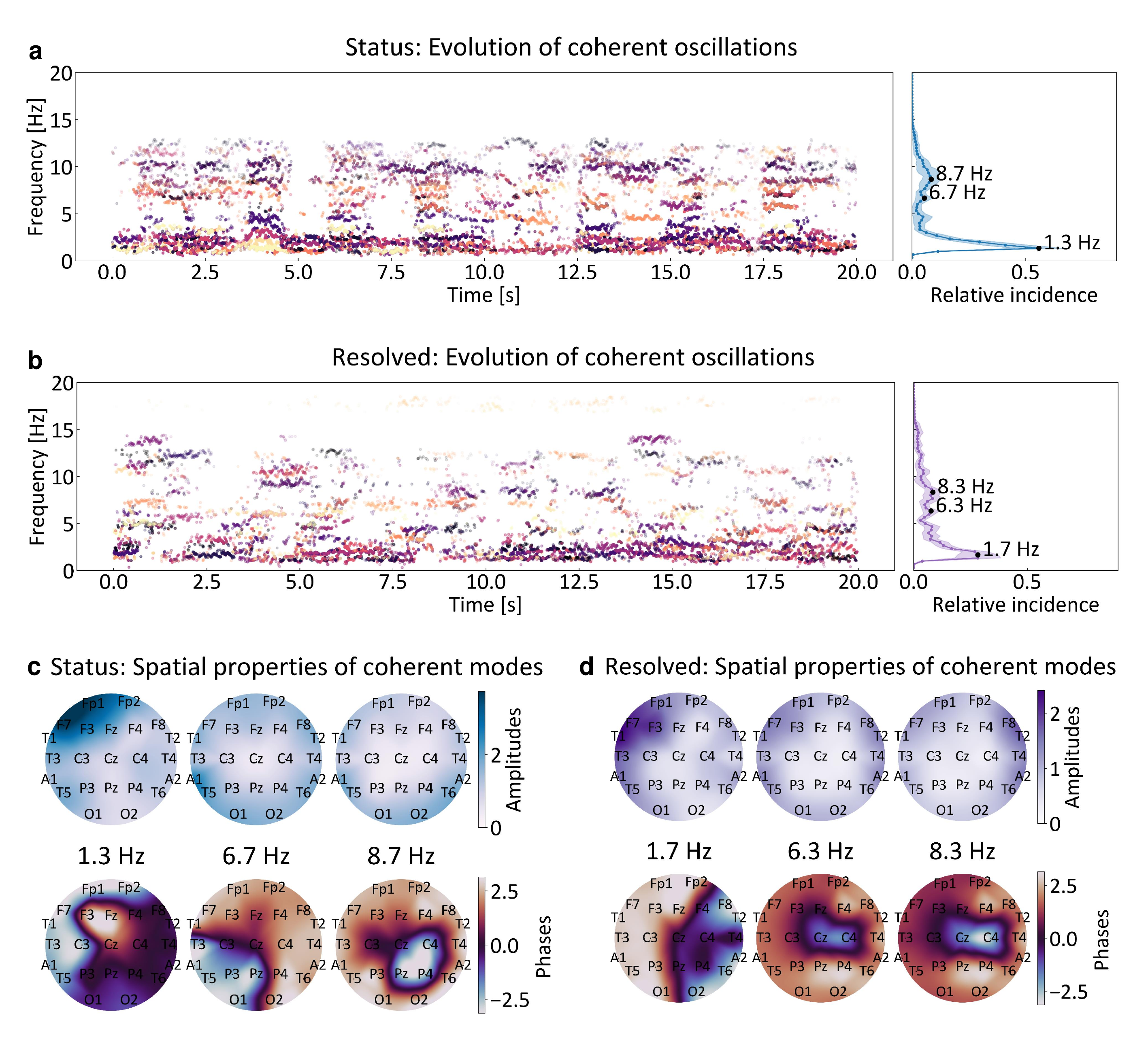}
\caption{
Robust coherent oscillations in the epileptic network |
By identifying the most persistently coherent oscillations directly from the recorded EEG data, our method discriminates the fundamental brain oscillations from accidental noisy contributions to the power spectrum.
\textbf{a}~Example of time evolution of the robust coherent oscillations in one $20$ s long epoch of status epilepticus recordings (left), and the normalized frequency distribution of their incidence in all considered epochs (right) for patient labeled as EEG 3. The different colors in the time evolution correspond to different coherent oscillatory modes, whereas the opaqueness is related to their strength. The peaks in the incidence distribution indicate frequencies at which pronounced coherent oscillations in the epileptic network are present.
\textbf{b}~Analogous results for the recordings after status epilepticus is resolved. The relative incidence of very low frequency coherent oscillations is reduced, and several peaks at higher frequencies appear.
\textbf{c},~\textbf{d}~Spatial profiles (distributions of amplitudes and phases) of the identified most prominent coherent oscillations for both states. The amplitude distributions indicate the EEG locations at which the strong coherent dynamics is most prominent. The phase distributions inform whether the coherent oscillations in different brain regions are fully in phase (homogeneous color) or a consistent phase shift is present (change of color).
}
\label{Fig1}
\end{figure}

Power spectral methods give insights into the power frequency dependence of certain dynamics, but do not inform about how coherent that power is. Coherence properties, on the other hand, are of great importance when considering the dynamics of the epileptic network and the oscillations therein, motivating the development of numerous coherence-related and interaction measures to include in EEG analyses~\cite{cliffUnifyingPairwiseInteractions2023}. However, extracting an interpretable model of dynamics from measures/metrics of power and coherence has remained challenging. We, therefore, chose to depart from this standard approach: We aimed for an approach that captures, on an individual level, all of the insights and characteristics that are contained in standard spectral and coherence measures, but, crucially, at the same time provides a clear path towards a comprehensive dynamical model of the underlying network that could lead to guiding individualized brain stimulation interventions.

The method we developed identifies the most persistent coherent oscillations directly from EEG data recorded at a certain point in time (Fig.~\ref{Fig0}\textbf{a}). It is based on a combination of an extended dynamical mode decomposition (DMD) algorithm~\cite{bruntonModernKoopmanTheory2021,bruntonExtractingSpatialtemporalCoherent2016}, an instance of the Koopman operator theory~\cite{bruntonModernKoopmanTheory2021}, and the theory of adiabatic evolution~\cite{joyeQuantumAdiabaticEvolution1994}. Coherent oscillations, automatically identified by the DMD algorithm, are followed through time via a sliding window, and only oscillations whose frequency and spatial profiles are stable, are kept as relevant (for more details about the basic algorithms and reasoning, see~\ref{SS_modeling_Koopman}~and~\ref{SS_modeling_adiabatic}). In contrast to PSD estimates, the method discriminates between robust coherent oscillations in the EEG dynamics and accidental noisy contributions to its power spectrum, so that fundamental brain oscillations can be recognized (Fig.~\ref{Fig0}\textbf{b}). Their properties and dynamical interactions are obtained in a fully data-driven manner and can serve for building up a patient-specific interpretable model of the core EEG dynamics (Fig.~\ref{Fig0}\textbf{c}), and be used to model its interaction with an external stimulation source (Fig.~\ref{Fig0}\textbf{d}).

We extracted the time evolution of the robust coherent oscillations for each patient, and each brain state (Fig.~\ref{Fig1}\textbf{a},\textbf{b}, Supplementary Information). The frequency distributions of the relative incidence of coherent oscillations in all epochs (Fig.~\ref{Fig1}\textbf{a},\textbf{b}, right panels) were used to determine the frequencies at which pronounced coherent oscillations in the epileptic network are present. As intuitively expected in status epilepticus, characterized by pronounced slow activity and epileptiform discharges~\cite{trinkaWhichEEGPatterns2015a}, the relative incidence of coherent modes had a significantly higher peak at very low frequencies ($\delta$-band). Smaller local distribution maxima at frequencies higher than the $\alpha$-band appeared only after the pathological state was resolved.

After determining the frequency of the pronounced coherent oscillations, we analyzed their spatial profiles, i.e., the corresponding distribution of amplitudes and phases on the scalp (Fig.~\ref{Fig1}\textbf{c},\textbf{d}). The high-amplitude regions in the amplitude distributions were used to detect regions in which the EEG dynamics consistently have strongly coherent oscillations. The phase distributions informed about the relative phase of the ongoing strong oscillatory dynamics. Namely, strongly coherent oscillations at a certain frequency can still have a consistent phase shift: in the topographic maps of phases, zero-phase synchronized dynamics are reflected in homogeneously colored scalp regions, while changes in color are related to permanent phase shifts (only a relative difference is meaningful in the context of phases). These phase shifts might be of crucial importance when trying to interact with the oscillating network by (resonant) stimulation.

\subsection{Generative coupled-oscillator model of network dynamics}
\label{SS_modeled}

Our next goal was constructing a corresponding effective individual model of brain dynamics (Fig.~\ref{Fig0}\textbf{c}). We aimed for a {\it personalized} model that can accurately reproduce the clinically relevant properties of the considered EEG dynamics: spectral content, inter-regional synchronization, and amplitude variation between different channels and states. In addition, we required a model that is: {\it interpretable}, in order to avoid black-box type of modeling and be able to combine the patient-specific models with existing knowledge about brain dynamics, and {\it predictive}, in order for the model to be useful when considering unseen scenarios, such as brain stimulation. All these criteria were fulfilled by a model that consists of coupled oscillators, algorithmically identified for each specific case via the extended DMD procedure (Fig.~\ref{Fig0}\textbf{c}).

\begin{figure}
\centering
\includegraphics[width=\linewidth]{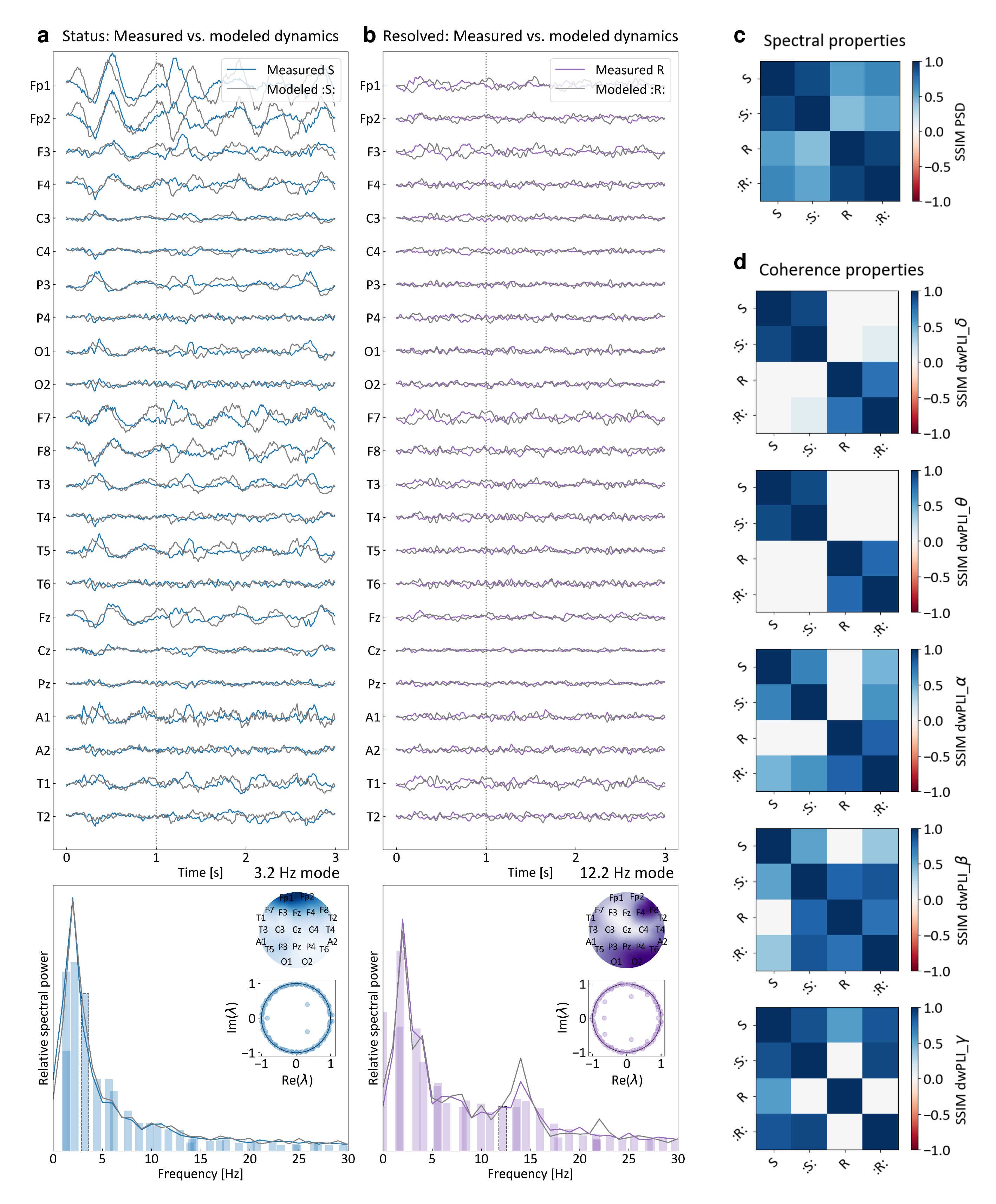}
\newpage    
\caption{Caption on next page.}
\label{Fig2}
\end{figure}

A non-overlapping sliding window was used to extract the time-evolution of the oscillator-network based model. For each time window, the generative model reproduced the EEG dynamics with the correct properties including spectral power, coherence, and amplitude (Fig.~\ref{Fig2}\textbf{a},\textbf{b}). Crucially, due to its oscillatory nature and the steady-state property of the data, the dynamics generated by the model fitted on a certain window could also be extrapolated to later times (Fig.~\ref{Fig2}\textbf{a},\textbf{b}, the fitting and extrapolation region are separated by vertical dashed lines). Precisely because of this property, the model could later be used to probe the network response under an external influence (Fig.~\ref{Fig0}\textbf{d}, Supplementary Information).

In order to systematically evaluate the accuracy and reliability of the EEG-driven models, we compared the standardized measures of spectral and coherence properties of the EEG dynamics, for all pairs of measured and modeled data in the two states. The spectral agreement was assessed by the time-averaged structural similarity index (SSIM) of power spectral density (PSD) (Fig.~\ref{Fig2}\textbf{c}, see~\ref{SSS_renormalized_SSIM}). An SSIM value of $1$ denotes perfect similarity, i.e. occurs when two identical properties are compared. The resulting SSIM matrix has a clear block-diagonal structure, with a very strong similarity ($\mathrm{SSIM} \mathrm{PSD} > 0.9$) within the same state, and a significantly weaker similarity ($\mathrm{SSIM} \mathrm{PSD} < 0.5$) between two different states. This holds for the measured data, the modeled data, as well as their mixed combinations. Intuitively, it means that the PSDs of the measured data from the two states have a certain degree of similarity, and this degree of similarity is very well reproduced by the generative model (approximately uniform color of the two off-diagonal blocks of the SSIM matrix in Fig.~\ref{Fig2}\textbf{c}).

The coherence comparison was assessed by the SSIM of the debiased weighted phase lag indices (dwPLIs), characterizing the degree of neural coherence in the five standard frequency bands
(Fig.~\ref{Fig2}\textbf{d}). The two lowest ($\delta$, $\theta$) have a clear block-diagonal structure of SSIM matrices (diagonal blocks $\mathrm{SSIM~dwPLI} > 0.9$, off-diagonal blocks $\mathrm{SSIM~dwPLI} < 0.1$). This is partly true for the $\alpha$-band too (diagonal blocks $\mathrm{SSIM~dwPLI} > 0.8$, off-diagonal blocks $\mathrm{SSIM~dwPLI} < 0.5$), and confirms that the model is selectively able to capture the lower-frequency EEG dynamics of the two states. The SSIM matrices for the dwPLI of higher-frequency bands ($\beta$, $\gamma$) lose the block-diagonal structure. Such a distinction between frequency bands is expected and understandable from different angles: (i) Neural oscillations at low frequencies are generally caused by synchronous changes in the membrane potentials of a large number of neurons~\cite{buzsakiScalingBrainSize2013}, thus being less sensitive to contributions of insignificant activity and being more coherent. This is also confirmed by the identified robust coherent oscillations, whose peaks rarely lie above $10$ Hz (Fig.~\ref{Fig1}). (ii) Since the model was established from the coherent oscillations in the EEG recordings, it does not emphasize high-frequency oscillations if these are not a reliable signature of the considered dynamics.

\subsection{Network response encoded in dynamical connectivity matrix}
\label{SS_responsiveness}

\addtocounter{figure}{-1}
\begin{figure}
\caption{
Coupled-oscillator model of epileptic network dynamics | 
The identified EEG-based coherent oscillations can be used to define a generative effective model of the brain dynamics. The data-driven model is constructed so to capture the clinically relevant properties of EEG dynamics: spectral content, inter-regional synchronization, amplitude variation between different channels and states.
\textbf{a},~\textbf{b}~Examples of compared measured and corresponding modeled EEG dynamics in the two states (top). The effective model was extracted from the first third of the shown dynamics (left of vertical dashed line), while its generative nature and steadiness of the modeled states were exploited to generate the rest. The frequencies and relative powers of the widespread oscillatory modes building up the model follow the corresponding spectral power distribution (bottom). Each of the modes has a defined spatial distribution. The inset plots show examples of such distributions (inset topo-plots) and stability exponents (real vs. imaginary components of eigenvalues $\lambda$) for one of the modes in each state (3.2 Hz in Status, as 12.2 Hz in Resolved).
\textbf{c} Time-averaged structural similarity indices (SSIM) of all combinations of measured and modeled power spectral density (PSD) in the two states (see legends in b,c). The block-diagonal structure of the resulting matrices clearly shows that the model is able to capture the differences as well as the similarities between the two states.
\textbf{d}~Analogous similarity comparison for the debiased weighted phase lag index (dwPLI) in each of the five epileptic bands. White fields denote pairs for which the comparison result was not statistically significant ($p>0.05$). The band dependent structure of the effective SSIM matrices (block-diagonal only for lower three bands) are a consequence of a significantly weaker coherence of higher frequencies brain dynamics.
}
\end{figure}

The dominant coherent oscillations extracted from the EEG recordings of each patient (Fig.~\ref{Fig1}, see~\ref{SS_coherent_oscillations}) indicate the frequencies, spatial profile and relative phases of stably synchronized neuronal firings, leading to meaningful endogenous brain oscillations. The response to a certain stimulus, however, also depends on the dynamic connections between the oscillatory elements in the network. Each personalized oscillator-network model (see~\ref{SS_modeled}) is encoded by a dynamical connectivity matrix $A$, which contains detailed information about both of these aspects: The eigenvalues and eigenvectors of $A$ encode the spatio-temporal properties of the coherent oscillations, while the matrix $A$ itself informs about how an instantaneous state of the network, $x_t$, containing the EEG potential value at all considered channels in the moment $t$, influences the instantaneous network state in the next moment,
\begin{equation}
\label{Eq_brain}
x_{t+1} = A x_t .
\end{equation}
The matrices $A$ fitted to the dynamics of EEG recordings are real matrices, but not necessarily symmetric. This is because two different brain regions, here represented by two EEG channels, do not always influence each other in the same way, i.e., do not project between each other symmetrically. 

Although the dynamical connectivity matrix contains extensive information about the dynamics of the system, it still remains hard to predict these dynamics by just considering the matrix elements. Nevertheless, Eq.~\ref{Eq_brain} is an example of a so-called equation of motion (EOM) for a dynamical system, which allows to predict the evolution at later times, as well as to probe the impact that external stimuli have. Given the oscillatory character of the models of interest, a natural form of stimulus to consider is time-periodic stimulation. Such reasoning is aligned with the standard principles on brain network dynamics, which have a pronounced oscillatory character and are influenced by time-periodic brain stimulation.

Therefore, we considered the response of the extracted patient-specific models of status epilepticus to time-periodic stimuli. We created a representative set of dynamical connectivity matrices $\{A_t\}$ for the pathological EEG dynamics of each patient, where each matrix $A_t$ corresponded to a different time window within the considered EEG recording. Different windows were non-overlapping, so to ensure their independence for later statistical analyses. The coupled-oscillator model defined by each of the matrices $A_t$ in the corresponding set was then probed by driving a certain node in the oscillator network with a weak sinusoidal stimulus of a certain frequency. Different stimulation locations and frequencies were tested. The promising choices of stimulation frequency were informed by the analyses of robust coherent oscillations (see~\ref{SS_coherent_oscillations}) and, more specifically, corresponded to the frequency peaks in the incidence distributions (Fig.~\ref{Fig1}\textbf{a}, right), which signaled the patient-specific frequencies at which the underlying brain dynamics have consistent oscillations. The resulting spatial distributions of the median strength of network response (quantified by the strength of the newly excited oscillations, as encoded in the network dynamical connectivity matrix, see~\ref{SS_modeling_stimulated}) for different personalized stimulation frequencies (see Supplementary Information) provide an advanced method for identifying the persistent endogenous brain oscillations and which brain regions are likely to respond to a chosen stimulus. Crucially, this analysis is based both on the local brain oscillation properties as well as the influences different brain regions exert on each other.

\subsection{A data-driven model of neural entrainment to periodic stimulation}
\label{SS_brain_entrainment}

After establishing a data-driven equation of motion (EOM) describing the dynamics in the epileptic network (Eq.~\ref{Eq_brain}), we considered whether such formulation can be extended to also capture the generally accepted mechanism and the signatures of online neural entrainment by periodic stimulation (Fig.~\ref{Fig0}\textbf{d}). At the current stage of research, neither the epileptic network dynamics nor the mechanisms through which periodic stimulation acts on it are fully clarified. Nevertheless, having a patient-specific model that correctly yields all the consented signatures would be valuable, as it would allow for the best quantitative and objective assessment of the differences between the available options of stimulation parameters (stimulation frequency and phase difference, stimulation location).

From a macroscopic dynamical perspective, the mechanism of interaction according to the neural entrainment theory, has the following properties~\cite{johnsonDosedependentEffectsTranscranial2020, liuImmediateNeurophysiologicalEffects2018, aliTranscranialAlternatingCurrent2013, schmidtEndogenousCorticalOscillations2014}: By entraining neurons around the stimulation location, the periodic stimulation causes an online increase in the oscillation power at the frequency of stimulation. The strength of the neuronal response is proportional to the stimulation amplitude. It is also highly dependent on the intrinsically preferred oscillation frequency of the local neurons, where it causes the strongest response. For this reason, tACS experiments often start by assessing the dominant individual oscillation frequency in the stimulated region, usually determined as the peak in the non-stimulated PSD in the band of interest~\cite{corcoranReliableAutomatedMethod2018a}. Finally, the stimulation-induced oscillations can spread through the epileptic network, depending on the network functional connectivity. 

Remarkably, all these dynamical properties hold for a network of oscillators that is driven by a periodic force: A single driven oscillator responds with an oscillation at the stimulation frequency~\cite{pikovskySynchronizationUniversalConcept2008}. The strength of the response is linearly proportional to the stimulation amplitude (see~\ref{Methods}), and depends on the difference between the stimulation frequency and the inherent oscillation frequency of the oscillator. In a network of oscillators, the driven oscillation spreads through the network couplings. The profile of the resultant network oscillation pattern (oscillation amplitudes and phases) is defined by the network dynamical matrix $A$, and obtained from the following equation of motion:
\begin{equation}
\label{Eq_brain_stimulation}
x_{t+1} = A x_t + \eta \sin(\omega_{d} t + \varphi_{d}) .
\end{equation}
Here $\eta$ is a vector that defines the amplitudes of drive on the network nodes (its dimensionality is equal to the dimensionality of $x_t$), $\omega_{d}$ is the driving frequency and $\varphi_{d}$ a vector containing the driving phases for all network nodes. The access to a data-driven matrix $A$ (see~\ref{SS_modeled}), therefore, opens the way to a dynamical model that explicitly exhibits all the known implications of periodic stimulation on the epileptic network dynamics via entrainment (Fig.~\ref{Fig0}\textbf{e}), yet in an objective and personalized manner.

To explicitly prove that the full model correctly captures the expected neuronal behavior under entrainment by periodic stimulation, we first systematically tested the model based on the EEG recordings of one of the patients in the cohort, with a designedly chosen set of stimulation parameters (Fig.~\ref{Fig3}{\textbf{a},\textbf{b}}). The changes in spectral properties under periodic stimulation, quantified by the corresponding PSD, confirmed all the expected signatures (Fig.~\ref{Fig3}\textbf{a}): entertainment at stimulation frequency (new peak in PSD), different response for different stimulation frequencies (stronger response for a stimulation frequency close to the inherent coherent oscillations in the epileptic network) and different stimulation locations (different spatial patterns), as well as the dependence of response on the stimulation amplitude. The changes in coherence properties under periodic stimulation, quantified by the changes of dwPLI (which allows to control for volume-conduction-like effects in our model), showed that coherence of a certain frequency band can be selectively targeted and altered by appropriately choosing the stimulation frequency (Fig.~\ref{Fig3}\textbf{b}). Namely, the strongest change of dwPLI (up to $|\Delta \mathrm{dwPLI}| \approx 0.2$) occurred in the bands which the stimulation frequency belonged to, while the majority of the $\Delta \mathrm{dwPLI}$ matrix elements for other bands were negligible ($|\Delta \mathrm{dwPLI}| \lesssim 0.05$). The predominant sign of each $\Delta \mathrm{dwPLI}$ matrix was determined by the relative phase of stimulation with respect to the endogenous brain oscillations at that frequency. 

\subsection{Model-based prediction of dynamics modulation in status epilepticus}
\label{SS_prediction_status}

Finally, we demonstrate how the identified network of coherent oscillations (Fig.~\ref{Fig0}\textbf{b}) and the data-driven models of entrained neural dynamics (Fig.~\ref{Fig0}\textbf{e}) could be used for a personalized brain stimulation therapy prediction in status epilepticus (Fig.~\ref{Fig0}\textbf{f}). 

\begin{figure}
\centering
\includegraphics[width=\linewidth]{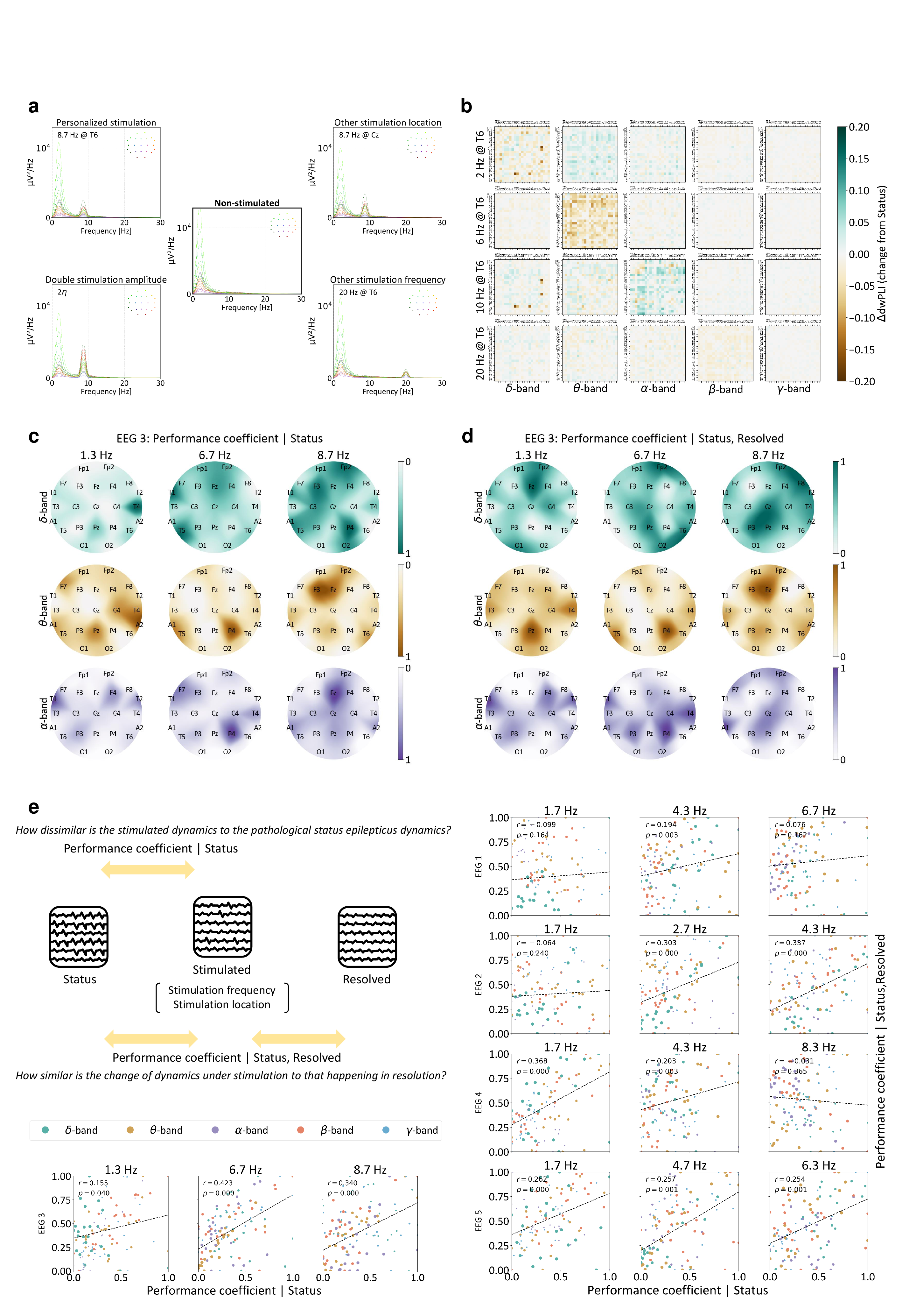}
\newpage    
\caption{Caption on next page.}
\label{Fig3}
\end{figure}

For each patient, the stimulation frequencies were chosen based on the dominant coherent oscillations (see~\ref{SS_coherent_oscillations}), which corresponds, under our assumptions, to a truly personalized dynamical investigation of the persistent intrinsic neural dynamics in the underlying epileptic network. Periodic stimulation (at one of those frequencies) of different brain regions was compared by considering the effects that stimulation of a certain node (channel) had on network coherence, thus also going beyond spectral properties (Fig.~\ref{Fig3}\textbf{c},\textbf{d}).

\addtocounter{figure}{-1}
\begin{figure}
\caption{
Epileptic network dynamics under periodic stimulation |
\textbf{a}~Model-predicted spectral properties of epileptic network under periodic stimulation, quantified by power spectral density (PSD). Power around the stimulation frequency increases, depending on the stimulation location and the intrinsic dynamical properties of the network. The change in power is also dependent on the stimulation amplitude, as observed in tACS experiments.
\textbf{b}~Model-predicted changes in coherence upon periodic stimulation in different frequency bands, quantified by debiased weighted phase lag index (dwPLI) per band. Phase synchronization is mostly altered in the stimulation frequency band.
\textbf{c}~Structural similarity (SSIM) of band dwPLI matrices between modeled stimulated dynamics and status epilepticus dynamics for EEG $3$. Stimulation frequencies are chosen based on the identified personalized dominant coherent oscillations. Low similarity for a certain choice of stimulation frequency and location is favorable.
\textbf{d}~Structural similarity (SSIM) of the model-predicted changes in band dwPLI upon periodic stimulation and upon status resolution. High similarity of change for a certain choice of stimulation frequency and location is favorable.
\textbf{e}~Weighted Kendall rank coefficients, quantifying the correlation between the performance coefficients based on only status epilepticus dynamics (c) and the performance coefficients obtained after knowing the way pathological brain activity is resolved (d). The p values of the weighted correlation coefficients in each subplot are obtained non-parametrically via bootstrapping.
}
\end{figure}

We first examined the extent to which periodic stimulation at different locations can change the pathological oversynchronized dynamics in status epilepticus. The departure from the pathological state was quantified by evaluating the renormalized similarities (see~\ref{SSS_renormalized_SSIM}), obtained from the SSIM index of the frequency band dwPLI matrices, between the modeled stimulated dynamics and status epilepticus dynamics (Fig.~\ref{Fig3}\textbf{c}). A low similarity of the stimulated and pathological state was conjectured as evidence for a successful departure from the pathological dynamics via periodic stimulation. To assess whether this premise holds, we also analysed the coherence changes happening upon periodic stimulation and compared them, by evaluating the renormalized change similarities (see~\ref{SSS_renormalized_SSIM}), to those happening after status epilepticus resolution (i.e., the "healthy" state) (Fig.~\ref{Fig3}\textbf{d}). Here, a high similarity of the dynamical change upon periodic stimulation meant that the stimulation is able to modulate the network dynamics in the way needed for a successful resolution of the pathological state, i.e., to drive the brain network closer towards the resolved (non pathological) state.

In a realistic scenario of a patient in status epilepticus, however, one would not have access to the resolved case when designing therapy by stimulation and choosing the stimulation parameters. Therefore, we systematically analysed whether the performance coefficient based on only status epilepticus dynamics (Status) correlates with the performance coefficient obtained after knowing the way the pathological brain activity is resolved (Status, Resolved), in all the patients from our cohort that experienced full pharmacological resolution (Fig.~\ref{Fig3}\textbf{e}). The correlation was quantified by the weighted Kendall rank coefficient, normally weighted based on the distance of the stimulation frequency to the center of each band, thus accounting for the selective spectral influence of entrainment via periodic stimulation (see~\ref{SSS_weighted_Kendall}). The statistical significance of these results was calculated via bootstrapping (see~\ref{SSS_weighted_Kendall}). The statistically significant ($p>0.05$) weighted correlations of the two types of performance coefficients are positive for all patients and all stimulation frequencies (Fig.~\ref{Fig3}\textbf{e}). The detailed behavior of the correlations is patient-dependent, emphasizing once again the high diversity in epileptic network dynamics, but also the value of access to clean EEG recordings in modeling and analysis: cleaner recordings (e.g., EEG 3) lead to better fitting models of the brain dynamics, thus facilitating the further steps of investigation. Nevertheless, the predominance of positive statistically significant correlations between the Status-based and (Status,~Resolved)-based performance coefficients, suggests that EEG-driven models of the epileptic network dynamics in status epilepticus alone can help predict the stimulation parameters that will lead to the resolution of this pathological brain state.


\section{Discussion}
\label{Discussion}

The heterogeneity of epilepsy, in terms of semiology, etiology, phenotypes, and genotypes, is remarkable. Many of the differences, however, have been overlooked within standard clinical and research approaches to diagnosis and treatment. Only recently, "personalized medicine" slowly started entering the epilepsy field, and introduced data-driven tailoring of pharmacological and surgical treatments~\cite{jirsaVirtualEpilepticPatient2017, nabboutImpactPredictivePreventive2020c, steriadeEnteringEraPersonalized2022, jehiMachineLearningPrecision2023, lhatooBigDataEpilepsy2020a}.
Nevertheless, similarly as in other domains of technology and society taking advantage of artificial intelligence  (AI)~\cite{adadiPeekingBlackBoxSurvey2018}, a key obstruction to the use of the diverse data-driven approaches in epilepsy is their lack of transparency and/or interpretability. 

Epilepsy patients are a vulnerable population, and as such underline the need for a careful approach to new methods and technologies. Yet, at the same time, due to the high percentage of uncontrolled epilepsies~\cite{devinskyEpilepsy2018}, it is also clear that these new technologies and methods are greatly needed. One, therefore, aims for approaches that unify data-driven individuality and personalized quantitative prediction, with interpretability in the context of standard clinical understandings and theories. In such an endeavour, we showed how the individual epileptic network dynamics and the underlying coherent neural oscillations can be captured by personalized generative EEG-driven oscillator-network models that are aligned with the existing neuroscientific intuition about brain oscillations and their modulation.

\textbf{Developing interpretable data-driven models of EEG dynamics}

The first contribution of this work is a new pathway for bridging the gap between interpretable models of brain-network dynamics and detailed personalized analyses of EEG recordings. While the latter can accurately quantify the spectral and coherence properties of network dynamics~\cite{dasSurveyEEGData2023}, they do not provide an intuition about the responsible dynamical mechanisms. On the other hand, models and theories of brain dynamics often revolve around neural oscillations~\cite{keitelRhythmsCognitionEvidence2022, friesRhythmsCognitionCommunication2015, engelDynamicPredictionsOscillations2001}, thus offering a link to the underlying dynamical mechanisms. Yet, they rely on generalized assumptions about the brain network and mostly neglect patient- or state-specific features of dynamics, such as an inherent stability of neural oscillations rather than simply their occurrence in a power spectrum. 

By identifying the most persistent coherent neural oscillations in the epileptic network and analyzing their evolution, we recovered key insights and characteristics contained in standard spectral and coherence measures. At the same time, the proposed approach also provided a direct personalized path for building generative dynamical models of the underlying network dynamics. Being fully data-driven and based directly on EEG recordings, the algorithm allowed us to fully circumvent generalized assumptions about anatomical and functional connections, and retain an objective quantitative picture of the network dynamics. By algorithmically discriminating between robust coherent oscillations and accidental noisy contributions to the power spectrum, we were able to recognize the patient- and state-specific fundamental brain oscillations.

The corresponding personalized models accurately reproduced the clinically relevant properties of EEG dynamics, including the spectral context, inter-regional synchronization and amplitude variation between different channels and states. By yielding an effective equation of motion, encoded by a dynamical connectivity matrix, the models opened the way to having digital twins of the epileptic network and predictions of unseen scenarios, e.g. the influence of external stimuli. In contrast to the majority of data-driven models, however, the obtained personalized generative models could also be understood in the context of existing neural-oscillation theories, due to the unique combination of the data-driven and interpretable Koopman operator theory.

Koopman operator theory~\cite{bruntonModernKoopmanTheory2021}, the basis which our modeling approach was built on, yields a high-dimensional linear representation of a non-linear dynamical system. The advantage of having a linear representation is the ability to rely on all the existing intuition and techniques for prediction, estimation and control that have been successfully developed for linear systems. The linearity of the representation, however, also leads to challenges. The epileptic-network dynamics is often strongly nonlinear, and is not necessarily always easily mapped to linear representations. While steady-state types of nonlinear dynamics, such as the one encountered in status epilepticus~\cite{burmanTransitionStatusEpilepticus2020} can be well mapped to linear representations of reasonable dimensionality, transient nonlinear dynamics, such as the one encountered in certain epileptic seizures~\cite{houssainiEpileptorModelSystematic2020} tends to require extremely high-dimensional linear representations, which can obfuscate the intuition around the underlying model. In addition, capturing the inherently nonlinear neuronal interactions with linear representations requires more intricate models when additional neuronal mechanisms are included, such as neural entrainment to external stimuli. The always-present noise in the analysed signals does not make such attempts easier, as it distorts the steady-state dynamics. This caused our analysis and model extraction to go beyond the standard textbook algorithms (requirement for adiabatic evolution of the coherent oscillations, effective models represented by sets of dynamical connectivity matrices). Nevertheless, the gain due to the combined data-driven and interpretable approach has been proven in numerous other fields that deal with realistic nonlinear dynamics, and we have here seen it in the context of the epileptic network as well. Moreover, as extensions of the original algorithms are introduced~\cite{bruntonModernKoopmanTheory2021, bruntonDiscoveringGoverningEquations2016, kalurDatadrivenClosuresDynamic}, it is becoming feasible to develop data-driven interpretable models that explicitly include parts of the underlying nonlinear manifolds of the epileptic network.

\textbf{Clarifying the dynamics of neural entrainment}

The second contribution of this work is the establishment of a direct correspondence between periodically driven models of brain-network dynamics\textemdash obtained directly from EEG recordings and able to fully capture the personalized dynamical characteristics, and the mechanism of neural entrainment\textemdash the main theory behind the influence that periodic stimulation has on neural dynamics.

The neural entrainment theory describes the online effects that external periodic stimulation, such as tACS, has on neurons in the brain network. The online effects are then considered to mediate the long-term effects~\cite{vogetiEntrainmentSpikeTimingDependent2022}. It is mostly based on microscopic single-neuron concepts and measurements~\cite{johnsonDosedependentEffectsTranscranial2020}, and extended to the macroscopic full network context in behavioral EEG experiments~\cite{polaniaStudyingModifyingBrain2018}. In this transition, however, many relevant concepts from the microscopic domain are lost\textemdash for a good reason. Namely, there is currently no full understanding of the link between the single-neuronal dynamics (neuronal spiking), as recorded by single-neuron measurements, and the dynamics of large neuronal populations (neuronal oscillations), as recorded by EEG. Predicting the behavior of a system as a whole (full brain network) from that of its individual parts (single neurons) is generally considered a nontrivial endeavour, due to the often arising unexpected emergent phenomena~\cite{artimeOriginLifePandemics2022}. Combining it with additional nonlinear interactions with the environment, causes further complications. Therefore, experiments that study the effects of entrainment on the EEG level, mostly neglect the microscopic picture, base the protocols on generalized assumptions, and quantify the long-term behavioral effects of periodic stimulation~\cite{polaniaStudyingModifyingBrain2018}. We attempted to bridge this gap by considerations that emphasize the dynamical properties in both domains, constructing an EEG-driven model of the epileptic network that explicitly captures all consented signatures of neural entrainment. There are two challenges that we faced: First, as a consequence of the linear representation of the brain network dynamics, building the interaction model required us to find the effective linear representation of neuronal entrainment. Linearity in the context of driven oscillations is accompanied by the relativity of stimulation amplitudes, which we had to account for when considering the modeled response of neuronal populations. Second, we had to find a mathematical formulation in the macroscopic domain that best mirrors the known effects from the microscopic domain.

Although accompanied by certain limitations, as discussed above, the introduced linear representation of the epileptic network indicated several important points, that have been touched upon in literature: The extent of entrainment depends on the inherent neuronal properties and the previously present oscillations\cite{corcoranReliableAutomatedMethod2018a}\textemdash peaks in spectral power are a possible choice for defining the personalized stimulation frequency, but do not necessarily differentiate coherent neuronal dynamics. The phase of the periodic stimulation with respect to the inherent neuronal oscillations matters\cite{krauseTranscranialElectricalStimulation2023a}\textemdash and can both enhance and reduce the ongoing oscillatory dynamics. Finally, having a good generative model of the brain dynamics opens the way to the best quantitative and objective assessment of new therapies.

\textbf{Predicting the performance of periodic brain stimulation therapies}

The third contribution of this work is the proposal for a comprehensive model-driven predictive analysis of the therapeutic performance that can be achieved by periodically stimulating a pathological state in the epileptic network\textemdash more precisely, the resolution of status epilepticus. The healthy brain dynamics and the dynamics in status epilepticus differ in several aspects, affecting both the spectral and the coherence domain. Therefore, multiple angles had to be considered when analyzing the effects that therapies have on the dynamics. 

First, by extracting the evolution and spectral properties of dominant coherent oscillations in the epileptic network, we addressed the angle of personalized inherent brain dynamics and starting choice of relevant stimulation frequencies. 

Second, by testing the model response to different stimulation frequencies, we addressed the spectral properties. In this context, there is a substantial difference between the lowest stimulation frequencies (corresponding to the peak of oversynchronized pathological activity in the $\delta$-band, and the higher stimulation frequencies (mostly lying within the $\theta$- and $\alpha$-bands). On the one hand, brain dynamics in status epilepticus have an overly pronounced $\delta$ power~\cite{trinkaWhichEEGPatterns2015a}, which makes the brain models highly responsive to low-frequency stimulation. The response, however, depending on the phase between the existing and entrained oscillations, can lead both to an increase and a decrease of this pathological band power\cite{krauseTranscranialElectricalStimulation2023a}\textemdash making low-frequency stimulation risky. On the other hand, the responsiveness of the brain models at higher frequencies is smaller, but it combines two potential benefits: an enhancement of band power in the "good" ($\theta$ and $\alpha$) bands, and a reduction of pathological band power, as some of the neurons are entrained to oscillate at a different rhythm. 

Third, by introducing the performance coefficients, we addressed the harder-to-grasp coherence domain of dynamics. The characterization of coherence changes is more complicated than that in the spectral domain, as it includes pair-wise relations and the corresponding changes thus happen in a higher-dimensional space: moving away from an instance of pathological dynamics does not necessarily imply moving in a favorable direction. We, therefore, introduced two metrics to quantify the performance of a certain choice of stimulation parameters: The renormalized coherence similarity to the status dynamics, quantifying the extent to which the chosen stimulation causes the dynamics to deflect from the pathological state, and the renormalized coherence change similarity, quantifying the extent to which the chosen stimulation causes the dynamics to change in the same direction as a successful pharmacological resolution would. The predominantly positive statistically significant correlations that we have shown between these two metrics suggest that, in a realistic status epilepticus scenario, where the design of the therapy by stimulation would have to be done without access to a resolved dynamics, EEG-driven models of the epileptic network dynamics in status epilepticus alone can suffice for choosing the best stimulation parameters.

\textbf{Outlook}

By aiming at interpretable EEG-driven models of the epileptic network that can predict the influence periodic stimulation has on it, we have faced two major outstanding challenges in the field of neuroscience.
First, the gap between microscopic and macroscopic theories and experiments: While neuroscientific research mostly concentrates on understanding single neuronal firings and their interactions, standard clinical approaches are based on EEG and other time-continuous macroscopic measurements. Due to the limited communication between these two fields, the micro-macro correspondence stays enigmatic in most cases.
Second, the gap between the nonlinearity of brain dynamics, and the linearity of human intuition and techniques for prediction: While it is known that many nonlinear mechanisms can be well captured by linear representations~\cite{bruntonModernKoopmanTheory2021}, these representations are often restricted to a particular regime and require more intricate approaches to modeling.

We have addressed both challenges by introducing effective models that account for the transition between single-neuronal and EEG findings, as well as the nonlinear effects arising in neural networks under stimulation. In the future, it would be beneficial to find an accurate and interpretable nonlinear representation of the epileptic network dynamics~\cite{bruntonModernKoopmanTheory2021, bruntonDiscoveringGoverningEquations2016, kalurDatadrivenClosuresDynamic}. More generally, studies to come should engage in understanding the mapping between microscopic neuronal spiking and the macroscopic EEG oscillations. We believe that the benefits of having an effective theory clarifying the basic mechanisms through which network oscillatory dynamics emerge from a system of many spiking elements would have a broad impact.

Finally, we would like to emphasize that the development of interpretable EEG-driven models of the epileptic network opens the way to personalized quantitative studies that go beyond low-frequency stimulation and tACS, both in the clinical as well the neuroscientific domain. As future extensions, one could assess the dynamical mechanisms that accompany the network modulation by chronic deep brain stimulation (DBS), develop personalized interpretable data-driven seizure models, or investigate sleep dynamics from a new quantitative model-based perspective.


\section{Methods}
\label{Methods}


\subsection{EEG recordings and preprocessing}
\label{SS_preprocessing}

We considered electroencephalographic (EEG) recordings of $10$ patients in non-convulsive status epilepticus. In $5$ of the cases, the status was resolved after the administration of pharmacological therapy. The EEG recordings were recorded on a clinical routine EEG device (\textit{Nihon Kohden}), with a 10-20 EEG system and a sampling rate of $200$ Hz. Since recordings of non-convulsive status epilepticus are minimally affected by artifacts, we only high-pass-filtered the EEG signals to $0.5$ Hz. Additionally, since we were mostly interested in the lower brain bands, we re-sampled the signals to $100$ Hz. This was motivated by the steep increase of the computational cost of modeling with the sampling rate. The local ethics committee approved the study, all patients or legal representatives gave informed consent. However, this consent did not include a provision stating that individual data can be made freely accessible.


\subsection{Modeling of network dynamics via Koopman theory}
\label{SS_modeling_Koopman}

The dynamics of the epileptic network, monitored through the EEG measurements, is an instance of a complex nonlinear dynamics. Owing to the complexity of the brain and the presence of various types of noise, it cannot by any means be represented through a simple set of equations. An optimal approach for modeling such a dynamics, thus, has to be significantly data-driven. If interested in a data-driven yet interpretable modeling paradigm for complex nonlinear systems, Koopman analysis emerges as an excellent basis.

\subsubsection{Koopman operator theory}
\label{SSS_Koopman}

Nonlinear dynamical systems are generally governed by the equation $\frac{d}{dt} x(t) = f(x(t), t, \beta, u(t))$,
where $x(t) \subseteq \mathbb{R}^n$ is the state of the system at time $t$, $f$ is a nonlinear vector field describing the dynamics, $\beta$ denotes equation parameters, and $u(t)$ is an external actuation. For an autonomous system that does not explicitly depend on time, the equation simplifies to
\begin{equation}
    \frac{d}{dt} x(t) = f(x(t)) .
\end{equation}
For the sake of clarity, we will from now on omit explicit additional dependence.
In practice, the state $x(t)$ is measured at discrete points in time ($x_t$) and governed by the corresponding discrete-time dynamical equation
\begin{equation}
\label{EOM_linear_discrete}
    x_{t+1} = F(x_t).
\end{equation}
An epileptic network, whose nonlinear dynamics is measured at a certain sampling rate (yielding a measurement $x_t$, whose dimensionality equals the number of EEG channels, for each sampling moment $t$), can be seen as such a system.
Following the Koopman operator theory~\cite{bruntonModernKoopmanTheory2021}, the nonlinear (but finite-dimensional) dynamics of a nonlinear system (Eq.~\ref{EOM_linear_discrete})
can always be expressed, through a suitable coordinate transformation $g$, as linear dynamics governed by the infinite-dimensional Koopman operator $\kappa$,
\begin{equation}
    g(x_{t+1}) = \kappa g(x_t) .
\end{equation}
The eigenvalue decomposition of the Koopman operator $\kappa$ yields the Koopman eigenfunctions $\varphi_\kappa(x)$ and corresponding eigenvalues $\lambda_\kappa$, which satisfy
\begin{equation}
    \varphi(x_{t+1}) =\kappa \varphi_\kappa(x_t) = \lambda_\kappa \varphi_\kappa(x_t) .
\end{equation}
The eigenfunctions $\varphi_\kappa(x)$ are time-invariant directions in the space of observables $g(x)$.
Specific linear combinations of the Koopman eigenfunctions give rise to time-invariant directions in the space of states $x$ directly, and are known as Koopman modes. Each of these modes captures a distinct spatial pattern in the behavior of the nonlinear system.

The goal is then to find a tractable finite-dimensional linear representation $K$ of the Koopman operator $\kappa$,
i.e. to find the effective coordinate transformation $g$, in which the nonlinear dynamics appears linear, 
$g(x_{t+1}) = K g(x_t)$.
By having such a linear representation of dynamics, one recovers all the intuition and mathematical framework that have been successfully developed for linear system, in which the principle of superposition is valid and the spectral decomposition (eigenvalues, eigenvectors) fully characterizes the dynamics. 
In the recent years, several computational algorithms for the identification of finite-dimensional representations of the Koopman operator $K$ have been developed, among which the most widely used is the dynamic mode decomposition (DMD).

\subsubsection{Dynamic Mode Decomposition (DMD)}
\label{SSS_DMD}

Dynamic Mode Decomposition (DMD) is one of the most standard and reliable algorithms to approximate the Koopman operator $\kappa$ from data~\cite{schmidDynamicModeDecomposition2010}. It seeks the best fit linear operator $A$ that approximately advances the state of the discrete-time system $x_t$ forward in time, to $x_{t+1}$,
\begin{equation}
    x_{t+1} = A x_{t} .
\end{equation}
The result of the DMD algorithm are the so called DMD modes, which can be interpreted as Koopman modes. DMD modes are related to the linear operator $A$ through the eigenvalue decomposition,
\begin{equation}
    A = \Phi \Lambda \Phi^\dagger,
\end{equation}
where $\Phi$ is the matrix of DMD eigenvectors, and $\Lambda$ the diagonal matrix of corresponding DMD eigenvalues.
The DMD modes correspond to spatially correlated structures that have some coherent linear behavior in time. The algorithm thus extracts the dominant spatially coherent oscillations, whose frequency and growth/decay rate are determined by the DMD eigenvalues, and whose spatial profile is determined by the DMD eigenvectors. 

\subsubsection{DMD algorithm}
\label{SSS_DMD_algorithm}

The DMD algorithm\cite{bruntonModernKoopmanTheory2021} is based on the computationally efficient singular value decomposition (SVD), thus making feasible its application to high-dimensional data. In addition, the procedure does not require using stochastic nor heuristic optimization (as often encountered in other machine-learning based modeling procedures), allowing to find solutions even when the high-dimensional landscape of solutions is complex and/or non-convex.

The algorithm starts by gathering several consecutively measured states $x_t$ and arranging as columns in the data matrices $X$ and $X'$,
\begin{align}
    X &= 
    \begin{bmatrix}
    | & | &    & | \\
    x_0 & x_{1} & \dots & x_{m-1} \\
    | & | &    & |
    \end{bmatrix}
    , \\
    X' &= 
    \begin{bmatrix}
    | & | &    & | \\
    x_{1} & x_{2} & \dots & x_{m} \\
    | & | &    & |
    \end{bmatrix}
    ,
\end{align}
which contain mostly overlapping data, time-shifted by one measurement.
These data matrices are then used to find the best fit linear operator $A$ through a high-dimensional linear regression of the dynamics that evolve $X$ to $X'$,
\begin{equation}
    X' = A X .
\end{equation}
A formal solution for the linear operator $A$ is given by 
\begin{equation}
    A \approx X'X^\dagger .
\end{equation}
The DMD algorithm does not compute $A$ first, but rather uses the singular value decomposition (SVD) of $X=U\Sigma V^*$, to directly find the its eigenvectors $\Phi$ and eigenvalues $\Lambda$\cite{bruntonModernKoopmanTheory2021}. This is particularly advantageous in cases of high-dimensional data (large $n$), as it avoids the prohibitively expensive eigenvalue decomposition of $A$. When the dynamics underlying the high-dimensional data has a low-dimensional structure, a good approximation can be achieved with relatively few DMD modes, characterized by smaller $\Phi$ and $\Lambda$ matrices, and obtained via the truncated SVD.
In any case, after obtaining the DMD modes, the observed dynamics $x_t$ can be modeled as
\begin{equation}
    {\hat x}_t = \Phi \Lambda^t \Phi^\dagger x_0.
\end{equation}

\subsubsection{DMD algorithm and EEG recordings}
\label{SSS_DMD_EEG}

The DMD algorithm was originally introduced in the fluid dynamics community. As such, it is meant for data that is extremely high-dimensional (large $n$), but not necessarily densely sampled through time\textemdash corresponding to a high spatial and low temporal resolution. Standard EEG recordings, on the other hand, have a relatively low spatial resolution, but high temporal resolution. That is, they are only about $25$-dimensional (the approximate number of recording channels for a 10-20 EEG setup), $n\sim 25$, but sampled at very high rates ($200$ Hz or more), $m\sim 200$ for a $1$-s time window.
The linear algebraic procedures in the algorithm, however, only make sense if certain criteria on the $X$ matrix dimensionality $n\times (m-1)$, encoded through the ratio of the spatial and temporal resolution, are respected. Namely, the maximal number of DMD modes that can be obtained is restricted by the smaller of $n$ and $m-1$. When $n$ is much smaller than $m$, there is a rank mismatch and the DMD modes that are obtained by the DMD algorithm are not enough to properly capture the system dynamics over the $m$ sampled states.
As a consequence, the standard DMD algorithm has to be adapted for electrophysiological recordings\cite{bruntonExtractingSpatialtemporalCoherent2016}: augmented data matrices have to be constructed, by vertically stacking $h$ time-shifted version of the original data matrices, which increases the number of channels to $hn$, e.g.,
\begin{align}
    X \rightarrow 
    \begin{bmatrix}
    | & | &    & | \\
    x_{0} & x_{1} & \dots & x_{m-1} \\
    | & | &    & | \\
    | & | &    & | \\
    x_{1} & x_{2} & \dots & x_{m} \\
    | & | &    & | \\
    & & \vdots \\
    | & | &    & | \\
    x_{h-1} & x_{h} & \dots & x_{m+h-2} \\
    | & | &    & | \\
    \end{bmatrix}
    .
\end{align}
The application of the DMD algorithm to the augmented data matrices yields $hn$-dimensional DMD modes, which roughly consist of $h$ copies of the non-augmented $n$-dimensional DMD modes. Intuitively, applying the DMD algorithm to the augmented data matrices is appropriate because data is sampled densely enough so that the time evolution at all time points closer than $h+m$ can be well approximated by the same non-augmented operator $A$ and the same non-augmented DMD modes. The goal is then to find the smaller stacking degree $h$ so that it provides enough DMD modes to capture the considered dynamics. For the specific choice of DMD hyperparameters ($n$, $m$, $h$), see~\ref{SS_hyperparameters}.


\subsection{Extraction of adiabatically evolving coherent oscillations}
\label{SS_modeling_adiabatic}

\subsubsection{Stable coherent oscillations from DMD modes}
\label{SSS_adiabatic_stable}

The DMD algorithm identifies the dominant coherent oscillatory modes in the time window defined by the considered data matrices. The length of these time windows is, however, restricted: Considering a very long time window means further increasing $n$, which then requires high stacking degrees $h$ in order to avoid a strong mismatch of the two matrix dimensions (see~\ref{SSS_DMD_EEG}). Having high stacking degrees in the data matrix, in return, implicitly assumes that the dynamics through a fairly long time is steady enough to be best approximated by the same choice of operator $A$\textemdash an assumption that is easily invalidated when considering EEG recordings. We, therefore, chose the minimal time window length $n$ that is enough to capture all the frequencies of interest, i.e., $n=200$ corresponding to $1$ s (see~\ref{SS_hyperparameters}). For each time window, we extracted the single-time-window ({\it instantaneous}) coherent modes by the DMD algorithm applied on the augmented data matrices.

The DMD modes inform about the dominant coherent oscillations in the epileptic network in one of the time windows of chosen length $n$. Not all of the identified DMD modes, however, necessarily reflect the oscillations that survive through longer time scales, as it can happen that a short accidental noise contribution dominates the stable brain oscillations in one time window. In order to identify the most persistent and stable coherent oscillations in the epileptic network, one needs to consider which of DMD coherent modes continuously arise as the DMD window slides through time. Therefore, we followed the evolution of the spatial and spectral profile of the DMD modes (determined by the DMD eigenvectors and eigenvalues, respectively), and only kept those modes that change slow enough, {\it adiabatically}. The evolution speed of eigenvalues for a stable system ($|\lambda|=|e^{i\omega}|=1$) can be quantified by the change in their frequency $\Delta f = f_{t+1}-f_{t} = (\omega_{t+1}-\omega_{t})/(2\pi)$, whereas the evolution speed of the complex eigenvectors is quantified by their Hermitian scalar product $o = \langle \varphi_{t+1}|\varphi_{t} \rangle$. 
For the choice of $\Delta f$ and $o$ hyperparameters, see see~\ref{SS_hyperparameters}.

\subsubsection{Effective eigenmodes via Gaussian kernel density estimation}
\label{SSS_adiabatic_modes_kde}

Once the most stable DMD modes were identified, we used them to define the effective long continuous oscillatory modes in the epileptic network. The effective frequencies of the dominant network oscillatory modes were signaled by the peaks of the frequency distributions (Fig.~\ref{Fig1}\textbf{a},\textbf{b}, right panels). The effective spatial profiles (amplitudes and phases) were obtained from the distributions of all identifies long continuous oscillatory modes. The spatial properties of each DMD mode are encoded in a complex eigenvector
\begin{align}
    \phi 
    &= \left[
    \phi^{(0)}, \phi^{(1)} \dots \phi^{(n_{ch}-1)}
    \right]^T \\
    &= \left[
    a^{(0)} e^{\varphi^{(0)}}, a^{(1)} e^{\varphi^{(1)}} \dots a^{(n_{ch}-1)} e^{\varphi^{(n_{ch}-1)}},
    \right]
\end{align}
where $n_{ch}$ is the total number of EEG channels, $\{a^{(j)}\}$ is the corresponding set of complex amplitudes and $\{\varphi^{(j)}\}$ is the corresponding set of complex phases.
We used Gaussian kernel density estimation (implemented by Python {\it stats} library) to non-parametrically estimate the joint density function of complex amplitudes $\{a^{(j)}\}$ and phases $\{\varphi^{(j)}\}$ for all stable coherent oscillations. The maxima of these density functions were then used to define the patient-specific effective amplitudes and phases of the dominant coherent oscillations. 


\subsection{Modeling epileptic network dynamics under periodic stimulation}
\label{SS_modeling_stimulated}

In this part we discuss the details and reasoning behind the introduced models of the epileptic network under the influence of periodic stimulation, as e.g., transcranial alternating current stimulation (tACS).

\subsubsection{Effective restriction of linear model response via spectral power}
\label{SSS_linear_power}

Following the entrainment theory~\cite{johnsonDosedependentEffectsTranscranial2020}, the effective interaction of the periodic stimulation and a neuronal network can be seen as a periodic drive of a network of linear oscillators (see see~\ref{SS_brain_entrainment}): the network response happens at the stimulation frequency, depends on the inherently preferred oscillation frequencies and the dynamical connectivity of the network, and is proportional to the stimulation amplitude. The value of the effective stimulation amplitude, however, is not absolutely defined and only ratios of amplitudes have a meaning: (i) Although certain parallels can be drawn by relying on heuristic analyses of the dominating dynamical properties, there is no exact straight-forward connection between the microscopic single-neuronal dynamics and the macroscopic network dynamics, nor a direct connection between the stimulation current amplitude and the effective strength which it stimulates the underlying network with. (ii) The response to a periodic drive of a (network of) linear oscillator(s) does not have an upper bound\textemdash the stronger the drive, the higher the response amplitude. This is in contrast with the actual biological situation, in which the amplitude of the oscillation is roughly proportional to the number of neurons that can be entrained to fire at the stimulation frequency and, therefore, inevitably bound from above.

Therefore, in order for the model to fully reproduce the experimental observations with tACS~\cite{johnsonDosedependentEffectsTranscranial2020}, a mechanism confining the response to biologically realistic values had to be incorporated in the interaction model. The exact percentage of neurons that can be entrained by tACS depends on the distribution of neuronal orientations with respect to the electric field, and the stimulation amplitude, but it was found to reach up to almost $\alpha_{max} \approx 30\%$~\cite{johnsonDosedependentEffectsTranscranial2020}. We implemented this aspect in the linear model by having the effective dynamics under periodic drive $x(t)$ as a combination of an unaltered activity $x^A(t)$ and a stimulation-entrained oscillatory activity $x^S(t)$,
\begin{equation}
    x_t = (1-\alpha) x^A_t + x^S_t .
\end{equation}
 The parameter $\alpha$ quantified the ratio of the neuronal population that was entrained for a certain choice of stimulation parameters (frequency, location, amplitude), and was measured by the signals standard deviations,
\begin{equation}
    \alpha \equiv \frac{\sigma(x^S)}{\sigma(x^A)}.
\end{equation}
The scale of the stimulation amplitudes $\eta_S$ was then chosen so to always remain in the biologically feasible range, $\alpha\leq \alpha_{max}$ on average. In this way, the effective dynamics was normalized, thus realistically corresponding to a response of a limited number of neurons.
In the implemented model, the unaltered dynamics $x^A(t)$ is followed by a $1-\alpha$ part of the total neuronal population and governed exclusively by the dynamical properties of the epileptic network, encoded in the dynamical connectivity matrix $A$,
\begin{equation}
    x^A_{t+1} = A x^A_t.
\end{equation}
The entrained dynamics $x^S(t)$ is followed by the remaining part of the total neuronal population and governed by an additional driving term in the equation of motion, corresponding to the periodic stimulation,
\begin{equation}
    x^S_{t+1} = A x^S_t + \eta_S \sin(\omega_S t + \phi_S) .
\end{equation}
Here $\eta_S$ is a vector containing the effective amplitudes of periodic driving in all the channels of the network, $\omega_S$ the stimulation frequency and $\phi_S$ the stimulation phase. Note that a stimulation with different phases for different channels can also be easily implemented, by promoting $\phi_S$ to a vector.

The above defined ratio of entrained neurons $\alpha$ was used to estimate the strength of the newly excited oscillations, encoded in the network dynamical connectivity matrix $A$\textemdash the epileptic network response for a certain choice of stimulation parameters.

\subsubsection{Stimulation phase}
\label{SSS_stimulation_phase}

The underlying brain dynamics in the same frequency range as the stimulation frequency is thought to have a large impact on how the unperturbed dynamics is influenced by the stimulation~\cite{krauseTranscranialElectricalStimulation2023a}. An important element in this context it the stimulation phase $\phi_S$ relative to the phase of the underlying brain oscillations. However, as long as a full closed loop stimulation paradigm is not implemented, which is highly nontrivial for frequencies that are not very low, the relative phase is not necessarily controlled in experiments. We, therefore, systematically chose a new random relative phase $\phi_S$ for each tested model in the representative set $\{A_t\}$.

\subsubsection{Stimulation amplitude}
\label{SSS_stimulation_amplitude}

As discussed above, the stimulation amplitudes $\eta_S$ do not have an absolute meaning, both because the relation between the exact stimulation current and the network drive is unknown, as well as because the response is unbounded in a network of linear oscillators. Therefore, two options are available. The first option is to fix the driving amplitude to a certain value, and then different combinations of other parameters (stimulation frequency, stimulation location) compared. This is the approach that we took in the analysis of network responsiveness to periodic stimulation. The second option is to fix the maximal response of the network dynamics, based on the biologically known maximal responsiveness $\alpha_{max}$, and then find the amplitude $\eta_S$ that leads to such a response. Due to the linear dependence of the response to the stimulation amplitude, $\alpha \sim \eta_S$, this amplitude can be determined by simply testing the response $\alpha^\mathrm{trial}$ with a trial $\eta_S^\mathrm{trial}$, and then using the equation \begin{equation}
    \eta_S = \frac{\alpha_{max}}{\alpha^\mathrm{trial}} \eta_S^\mathrm{trial} .
\end{equation}
We took this approach in the second part of analysis, where we explicitly examined the properties of the modeled stimulated dynamics.


\subsection{Quantitative and statistical tools}
\label{SS_quantitative_statistical}

\subsubsection{Renormalized structural similarity index and performance coefficients}
\label{SSS_renormalized_SSIM}
The structural similarity index (SSIM) estimates the similarity between two matrices, originally containing the pixels of images for which it was developed~\cite{wangImageQualityAssessment2004}. Unlike most assessments that are based on quantifying the difference for each element, the SSIM extracts three high-level features (luminance, contrast, structure). It has a value of $1$ for two identical matrices. Because of its balanced sensitivity to absolute and relative distances in the considered matrices, the SSIM was recently used in EEG analysis~\cite{ipinaModelingRegionalChanges2020, perlGenerativeEmbeddingsBrain2020}.
We used the SSIM to estimate the similarity between the computed power spectral densities (PSD) and debiased weighted phase lag indices (dwPLI). The statistical significance of the evaluated SSIM coefficients was obtained non-parametrically by comparison with a distribution of surrogate dwPLI matrices corresponding to randomly permuted channels.

In Fig.~\ref{Fig3}, we were interested in a comparison including different patients and different states (induces by different stimulation conditions). To make this comparison possible, we implemented a linear transformation (renormalization) on the calculated values of SSIM, so to achieve a span between $0$ and $1$ for each patient and condition, yielding the presented performance coefficients. Namely, for each patient, the performance coefficient was obtained as
\begin{equation}
    \text{Performance coefficient} = 
    \frac
    {\mathrm{SSIM} - 
    \min\left(\mathrm{\{SSIM\}}\right)}
    {\max\left(\mathrm{\{SSIM\}}\right) - \min\left(\mathrm{\{SSIM\}}\right)},
\end{equation}
where $\mathrm{\{SSIM\}}$ denotes the set of all SSIM indices for a certain patient, a certain choice of stimulation frequency $\omega_S$, and a certain frequency band for which the dwPLI is evaluated. 

\subsubsection{Weighted correlations of performance coefficients}
\label{SSS_weighted_Kendall}
The performance coefficients were used to asses the correlations between the Status-based and (Status, Resolved)-based predicted performances. The correlations were quantified by a weighted Kendall rank coefficient. The Kendall rank coefficient was chosen because it does not assume normal distributions (non-parametric), and is more robust for small datasets with potential outliers. The weighted version of the Kendall rank coefficient was used because of the expectation that stimulation at a certain frequency mostly affects the dynamics in the frequency band around it. The relative weights $w_b$, were set by the distance of the stimulation frequency $\omega_S$ from the center of the considered frequency band, with a Gaussian weight function whose center was determined by the band center and whose standard deviation was determined by the band width. Namely,
\begin{equation}
    w_b\left(\omega_S\right) \sim
    \exp^{-\frac{\left(\omega_S/(2\pi) - f_b\right)^2}
    {2 \sigma_b^2}},
\end{equation}
where $f_b$ is the central frequency of the considered band $b\in\{\delta,\theta,\alpha,\beta,\gamma\}$, and $\sigma_b$ is its total bandwidth.

The statistical significance for weighted Kendall rank correlation coefficients was evaluated by constructing bootstrapped distributions. For each considered patient and stimulation frequency (Fig.~\ref{Fig3}\textbf{e}), we calculated the actual weighted correlation coefficient $r$ between the Status-based and (Status, Resolved)-based performance coefficients, as well as a corresponding weighted bootstrapped distribution of correlation coefficients $\mathcal{P}_r$. The bootstrap distribution $\mathcal{P}_r^{(b)}$ for every frequency band $b$ was constructed based on the correlation coefficients $r^{(b)}$ obtained for a surrogate dataset. The surrogate dataset consisted of $9999$ samples, of which each was obtained by randomly sampling with replacement the original pairs of performance coefficients. The size of each sample was equal to the size of the original sample for which the actual weighted correlation coefficient $r$ was computed. 
The band-specific distributions $\mathcal{P}_r^{(b)}$ were then combined to a joint distribution through the band-specific weights,

\begin{equation}
    \mathcal{P}_r = \frac{\sum_b w_b \mathcal{P}_r^{(b)}}{\sum_b w_b} .
\end{equation}
The non-parametrically estimated p-value for the weighted Kendall rank correlation coefficient $r$ was equal to the ratio of values in the distribution $\mathcal{P}_r$ with an opposite sign than the actual correlation coefficient $r$,
\begin{equation}
    p =
    \begin{cases}
        \sum \mathcal{P}_r(r<0) / \sum \mathcal{P}_r, & r\geq 0\\
        \sum \mathcal{P}_r(r<0) / \sum \mathcal{P}_r, & r < 0 
    \end{cases}
    .
\end{equation}


\subsection{Modeling and analysis hyperparameters}
\label{SS_hyperparameters}

In this part we discuss the choices made for the various hyper-parameters of the model. 

\subsubsection{Preprocessing of raw signals}

The raw data was notch-filtered at $50$ Hz, to remove the line noise artifacts. In addition, a Savitzky-Golay filter of second order, with a sliding window of $1$ s, was applied to remove a constant offset and very slow non-biological oscillations.
Since the relevant coherent oscillations and related dynamical behavior only occur at frequencies that are never higher than $30$ Hz, we then down-sampled the signals to $100$ Hz. This significantly reduced the computational complexity in the later steps, thus speeding up the modeling procedure (most of the steps in our procedure have a higher than linear dependence of computational time on the sampling frequency).

\subsubsection{Epoching}

Data best representing both dynamical states (Status and Resolved) was stored in an equal number of $20$ s long epochs. Due to the big difference in the total recording times between different patients , this number is not fully uniform between individuals  (between $5$ and $15$ epochs, depending on the recording), but does not pose a problem as models are anyway individualized. Modeling was performed by sliding a $1$ s long window through all epochs (see~\ref{SSS_hyperparameters_DMD}).
The standard estimation of PSD and dwPLI was performed by epoching the same data to $5$ s long epochs. 

\subsubsection{Dynamic mode decomposition}
\label{SSS_hyperparameters_DMD}

The sliding window length was set to $1$ s, allowing to capture the biologically relevant frequencies ($>0.5$ Hz) and in agreement with other work based on linear models~\cite{liNeuralFragilityEEG2021}.
As discussed in see~\ref{SSS_DMD_EEG}, due to the difference between the temporal and spatial resolution in EEG recordings as compared to fluid dynamical systems for which DMD algorithms were developed, copies of single-time-step shifted data had to be stacked, before the standard DMD procedure could be applied (see~\ref{SSS_DMD_EEG}). The need for stacking is a consequence of the incommensurate two dimensions of the matrices containing $\sim 20$ spatial points (EEG electrodes) and $\sim 100$ time points, leading to strongly non-square matrices with a limited number of eigenvalues. We found that a stacking factor $h=10$ optimally captured the properties (spectral, coherence) of the modeled signals, which is in agreement with existing literature\cite{bruntonExtractingSpatialtemporalCoherent2016}.
We applied the DMD algorithm (see~\ref{SSS_DMD_algorithm} without truncation, again owing to the relatively low dimensionality of the EEG data. In addition, the requirement a slow adiabatic evolution in the next modeling steps (see~\ref{SS_modeling_adiabatic}) lead to a further dimensional reduction.
The norms of the eigenvalues obtained via the dynamic mode decomposition are a measure of dynamical stability. Since we modeled steady state dynamics of the epileptic network, small deviations from perfect stability were attributed to noise and the corresponding dynamical matrices were regularized to only have eigenvalues $1$ and $0$. The exact number of these two types of eigenvalues is determined by the choice of the stacking factor $h$, i.e., by the dimensions of the data matrix in the currently analyzed time window.

\subsubsection{Extraction of continuous coherent oscillations}

We performed the dynamical mode composition with parameters described above, and sliding windows that were shifted by the sampling period ($1/100$ s). The continuous coherent oscillations were then identified by finding oscillatory modes that are the most stable in time through (as described in ~\ref{SSS_adiabatic_stable}), starting at points shifted by $1$ s and followed through the whole length of an epoch. By only taking into account the $10$ strongest modes at each time point, we performed a dimensional reduction. The requirements that we set for an oscillation mode to be considered continuous and slowly evolving (see~\ref{SSS_adiabatic_stable}) were an eigenfrequency that did not fluctuate more than $\Delta f = 0.5$ Hz and a minimal eigenvector overlap of $o = 80$ \%. The longest oscillatory modes, i.e., oscillatory modes whose length exceeds the mean mode length, were kept for further analysis.

\subsubsection{Optimization of spectral properties in models}

Models of the epileptic network dynamics were extracted by performing DMD with parameters described above, and sliding windows that are shifted by the whole length of the window, leading to their statistical independence. This yielded a set of dynamical connectivity matrices $A$ that best represented the data. In addition, a Broyden–Fletcher–Goldfarb–Shanno (BFGS) algorithm was applied to find the boundary conditions that optimally reproduce the spectral properties of the original data. Before comparison, spectra were smoothed by Gaussian convolution with $\sigma=0.5$ Hz, in order to disregard accidental deviations and facilitate the optimization process. As the dynamical matrices inherently contain information about the different channels, optimizing the agreement between averaged spectra quickly lead to very similar results as those that would have been obtained by separately optimizing the agreement in each channel.



\section*{Acknowledgements}
We are grateful to Matija Varga, Giuseppe Zito and Dominik Haener for inspiring discussions. 
This work was funded by the Swiss National Science Foundation grant (SNF 197766, awarded to L.I and R.P.), by the European Research Council (ERC) under the European Union’s Horizon 2020 research and innovation program (grant agreement no. 758604), by a European Research Council (ERC) starting grant (ENTRAINER) and by an ETH Grant (ETH-25 18-2, awarded to R.P.), by the Koetser Foundation research grant (awarded to T.D and L.I.) and by the research grants from the Swiss Epilepsy Foundation.


\section*{Author contributions}
T.D., L.I. and R.P. conceived the project. 
T.D. developed the modeling paradigm, and performed the analytical and numerical calculations.
L.I and R.P. supervised the project. 
All authors contributed to the discussion of the results and manuscript writing.

\printbibliography


\newpage
\setcounter{section}{0} 

\title{SUPPLEMENTARY INFORMATION\\
Personalized identification, prediction, and stimulation of neural oscillations via data-driven models of epileptic network dynamics}

\maketitle

\section*{Network response encoded in dynamical connectivity matrix}

\begin{figure}[H]
\centering\includegraphics[width=1.0 \linewidth]{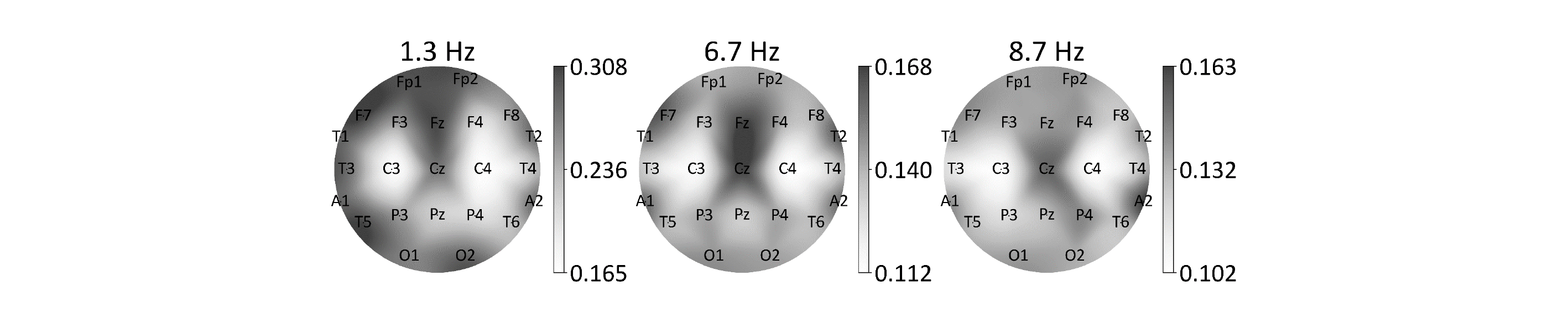}
\caption{Spatial distributions of the median strength of network response (quantified by the strength of newly excited oscillations, as encoded by the network dynamical connectivity matrix) for different personalized stimulation frequencies | EEG 3}
\label{SI_Fig3_responsiveness}
\end{figure}

After extracting the patient-specific models of Status dynamics, we explicitly tested their response to periodic stimulation. The promising choices of stimulation frequency corresponded to the frequency peaks in the frequency incidence distributions, which signaled the patient-specific frequencies at which the underlying brain dynamics have consistent oscillations. We computed the strength of the network response (quantified by the strength of the newly excited oscillations, as encoded in the network dynamical connectivity matrix) for each epoch (each $A_t$) corresponding to the considered patient and brain state. The effective strength of the network response was then calculated as the median of the results for all epochs (shown for EEG 3 in Fig.~\ref{SI_Fig3_responsiveness}).
See main text for further interpretation and discussion.

\section*{Robust coherent oscillations in the epileptic network}

\begin{figure}
\centering\includegraphics[width=1.0 \linewidth]{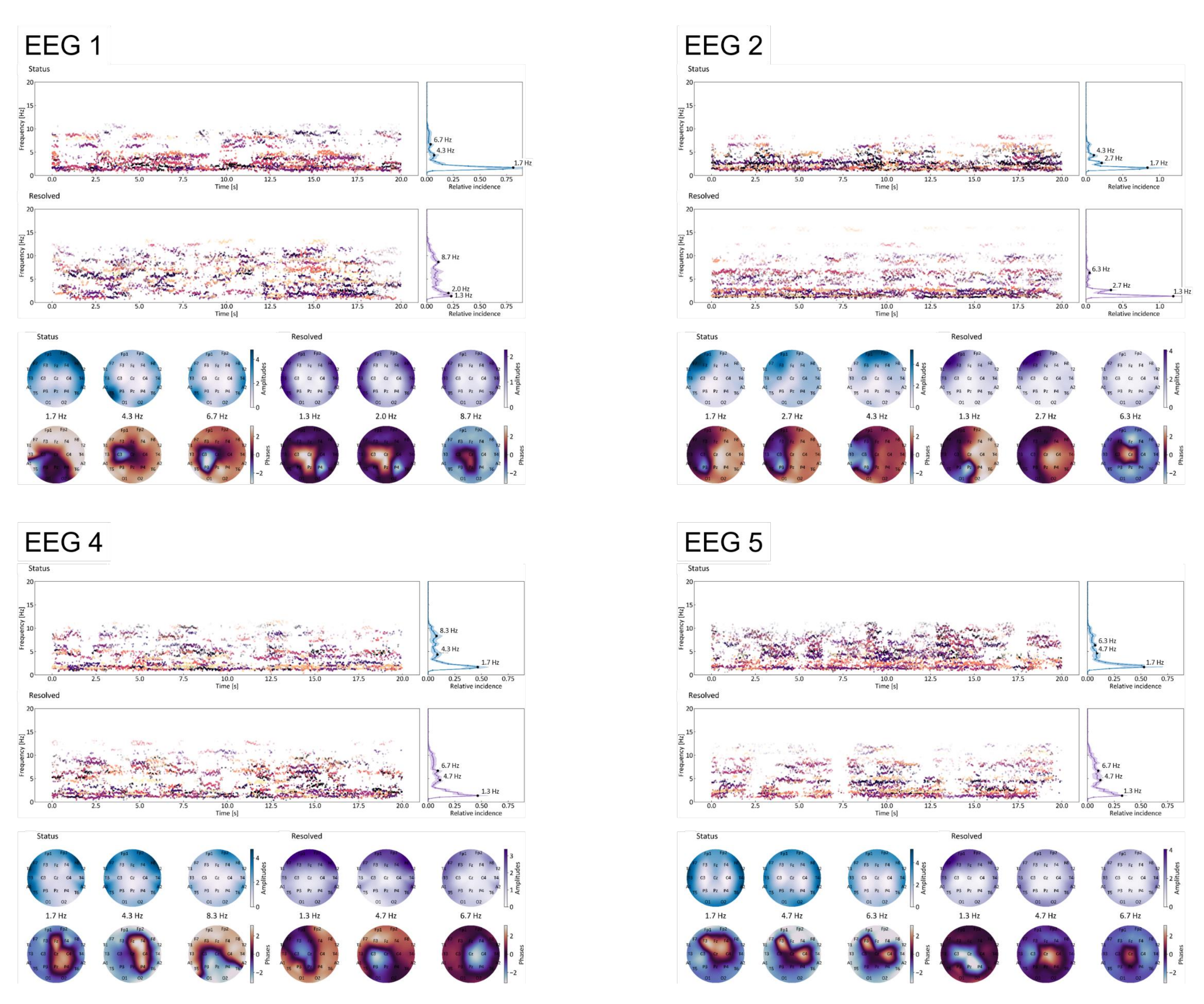}
\caption{Time evolution, frequency distributions and spatial profiles in Status and Resolved states for all patients not shown in the main text. | EEG 1,2,4,5}
\label{SI_Fig1_continuous_oscillations}
\end{figure}

We extracted the time evolution of the robust coherent oscillations for each patient with both the Status and Resolved states present. In the main text, we show the results for the exemplary patient EEG 3. In Fig.~\ref{SI_Fig1_continuous_oscillations} we show the results for all other patients. The plots are analogous. Through this procedure, we can identify the dominant coherent oscillations individually for each patient and brain state: the frequencies of the dominant oscillatory modes are indicated by the peaks in the frequency incidence distributions, whereas the topo-plots (amplitudes and phases of the oscillatory modes) characterize their spatial profile.
See main text for further interpretation and discussion.

\end{document}